**Spin-transfer torque switching in nanopillar superconducting-magnetic hybrid Josephson junctions[a]**

Burm Baek, William H. Rippard, Matthew R. Pufall, Samuel P. Benz, Stephen E. Russek, Horst Rogalla, and Paul D. Dresselhaus

National Institute of Standards and Technology, Boulder, CO 80305, USA

**The combination of superconducting and magnetic materials to create novel superconducting devices has been motivated by the discovery of Josephson critical current ($I_{cs}$) oscillations as a function of magnetic layer thickness and the demonstration of devices with switchable critical currents. However, none of the hybrid devices have shown any spintronic effects, such as spin-transfer torque, which are currently used in room-temperature magnetic devices, including spin-transfer torque random-access memory and spin-torque nano-oscillators. We have developed nanopillar Josephson junctions with a minimum feature size of 50 nm and magnetic barriers exhibiting magnetic pseudo-spin-valve behavior at 4 K. These devices allow current-induced magnetization switching that results in 20-fold changes in $I_{cs}$. The current-induced magnetic switching is consistent with spin-transfer torque models for room-temperature magnetic devices. Our work demonstrates that devices that combine superconducting and spintronic functions show promise for the development of a nanoscale, nonvolatile, cryogenic memory technology.**

Superconducting-magnetic hybrid devices[1,2,3,4,5,6,7,8,9,10] are being investigated as potential switching elements for low-energy cryogenic memory, which is essential for the realization of a high-performance energy-efficient superconducting computer[11,12,13,14,15,16,17,18,19]. A Josephson junction (JJ) incorporating a pseudo-spin-valve (PSV) barrier (a barrier containing two magnetic layers with different switching fields) is one of the simplest hybrid structures that allows switching of the superconducting critical current ($I_{cs}$) through control of the magnetic state[20,21,22]. Bell et al.[7] modulated $I_{cs}$ of such a device by changing the magnetization state of their PSV barrier. Recently, we showed that such a modulation can originate from either an exchange-field effect or a remanent-field effect and that the former may be used to build a nanoscale device in which digital information is stored as either Josephson energy or phase[10].

---

[a] Contribution of NIST, an agency of the U.S. government, not subject to U.S. copyright.

In a qualitative picture of superconductor-ferromagnet (S-F) physics, a Cooper-pair spin state evolves sinusoidally in the ferromagnetic barrier F of an S-F-S JJ, which results in a spatial modulation of the order parameter and an oscillation in $I_{cs}$ with magnetic layer thickness $d_F$, including sign changes with a period $\approx 2\pi\xi_F$ where $\xi_F$ is the characteristic oscillation length in F[4,23,24]. These sign changes indicate where the JJ switches the phase by $\pi$, called 0-$\pi$ transitions. This effect can be extended to a PSV barrier with two magnetic layers, F1 and F2 in which the oscillatory order parameter modulation is given by the different effective magnetic barrier thicknesses $x^P = x_{F1} + x_{F2}$ and $x^{AP} = x_{F1} - x_{F2}$, where $x_{Fi} \equiv d_{Fi}/\xi_{Fi}$ ($i = 1, 2$) for the parallel (P) and antiparallel (AP) magnetization states, respectively. Thus, by controlling the magnetization orientation of F2 relative to F1 (selecting either P or AP states), the Josephson coupling can be switched in amplitude ($I_{cs}$) or phase (0 or $\pi$) (Fig. 1a)[10].

Nanoscale JJs have not been extensively studied because the superconducting critical current density $J_{cs}$, of typical insulating or high resistivity barriers yields correspondingly small $I_{cs}$, which is difficult to measure. JJs with low-resistance metal barriers allow for a higher $J_{cs}$ at a cost of JJ speed (due to a longer single flux quantum pulse width $\approx \Phi_0/I_{cs}R_n$, where $\Phi_0$ is the magnetic flux quantum and $R_n$ is the normal state resistance). This may be an acceptable trade-off depending on the application. Room-temperature measurements of PSVs have shown that as the device size is reduced, current-induced magnetization switching (CIMS), based on the spin-transfer torque (STT) effect, is possible[25,26,27] and may be applicable to JJ systems. In a nanopillar PSV, electrons flowing from the reference layer to the free layer are spin-polarized and result in a torque on the free-layer moment that aligns the moment parallel to the reference layer. If the current is reversed, the free-layer moment can be aligned in the anti-parallel orientation. This STT effect is scalable because the switching current decreases with the device area[25,26]. Here, we developed nanopillar JJs with PSV barriers and found that the exchange-field effect on $J_{cs}$ persists to at least the 50 nm scale and allows for the differentiation between P and AP states with a significant change in $I_{cs}$ in the superconducting state. We demonstrate complete magnetization reversal by the STT effect by comparing it with field-induced magnetization switching (FIMS) in $Ni_{0.8}Fe_{0.2}$/Cu/Ni-based JJ devices and showing the same relative changes in $I_{cs}$ across multiple Ni/Cu/Ni-based devices through the scalable exchange-field effect on the superconducting order.

The device structure we investigated is $Si/SiO_2/Nb(100)/Cu(3)/PSV/Cu(3)/Nb(200)$, where the numbers in parentheses indicate layer thickness in nanometers. We deposited $Ni_{0.8}Fe_{0.2}(0.8$ or $1)/Cu(5)/Ni(1.2$ or $2.4)$ or $Ni(1)/Cu(5)/Ni(2.4)$ for the PSV. We fabricated JJ devices using common magnetic nanopillar fabrication processes to produce ellipses with dimensions ranging from 50 nm × 100 nm to 300 nm × 600 nm. Fig. 1b shows the schematic for our device, which were mounted in a cryogenic probe and measured in a liquid-helium bath at 4 K. We used a superconducting magnet to apply a magnetic field parallel to the major axes of the elliptical devices. $I_{cs}$ and $R_n$ were extracted from the measured $I$-$V$ curves by use of least-squares fits to the expected electrical characteristics of the resistively-shunted junction[28].

Our S-PSV-S JJ devices show $I$-$V$ characteristics with different $I_{cs}$s depending on the relative orientations of the magnetizations of the two magnetic layers (Fig. 1c). While the normal resistance $R_n$ in our PSVs changes by less than 1 % at ≈10 K as a result of the giant magnetoresistance effect, we achieve a dramatic 2000% change in $I_{cs}$ at 4 K due to our careful selection of materials and thicknesses to produce a very small critical current in the P state; in Fig. 1c the equivalent metric is $|\Delta I_{cs}|/I_{cs}^P \approx 2000$ %, where $\Delta I_{cs} \equiv I_{cs}^P - I_{cs}^{AP}$, with $I_{cs}^P \approx 0.4$ μA and $I_{cs}^{AP} \approx 10$ μA. The field required to saturate the PSV magnetization is higher than 200 mT but at around 100 mT the magnetic flux gets trapped in the device and complicates the subsequent zero-field characterization. To address this problem, we heat the chip just above the Nb superconducting transition temperature $T_{cs} \approx 9$ K after applying the magnetic field pulse that sets the PSV state, then cool to 4 K in zero field before measuring each $I_{cs}$. We varied the field pulse height and obtained hysteretic changes in $I_{cs}$ resulting from the different magnetization switching fields of the two magnetic layers. Fig. 1d shows that the lower-coercivity layer $Ni_{0.8}Fe_{0.2}$ switches at $\approx 5$ mT (resulting in an increase in $I_{cs}$) and Ni switches over a field range from 40 mT to 120 mT (resulting in a decrease in $I_{cs}$). Separately, we measured the coercivities from magnetization loops of unpatterned $Ni_{0.8}Fe_{0.2}$ and Ni films and obtained 1 mT and 40 mT, respectively. In a lower field range (below the Ni switching fields), we could control the $Ni_{0.8}Fe_{0.2}$ magnetization direction without flux-trapping to obtain high and low $I_{cs}$ states associated with AP and P states, respectively, as shown in Fig. 1e.

Figs. 1f and 1g show that different results can be obtained with different fixed layer thicknesses. The opposite signs in $\Delta I_{cs}$ result from the oscillatory $I_{cs}$ vs. magnetic layer thickness characteristics. If the slopes in $I_{cs}$ vs. $d_{Ni}$ are opposite to each other (e.g., the two regions marked by the solid and dashed blue curves in Fig. 1a), the same change in effective magnetic thickness from P-to-AP switching can result in opposite signs in $\Delta I_{cs}$. The opposite signs of $\Delta I_{cs}$ with $d_{Ni} = 1.2$ nm and 2.4 nm are consistent with the results obtained in Ref. 10. The curvature in the data for a 300 nm × 600 nm elliptical device (Fig. 1g) is a part of the common Fraunhofer-like $I_{cs}$ response to the applied fields[28]. Although this effect, when combined with the remanent fields in the magnetic barrier, could result in a significant modulation in the maximum supercurrent at a zero applied field in a large JJ[10,29], this effect is not significant in our nanopillar devices due to the broad Fraunhofer-like patterns and the dominant behavior of the exchange-field effect[10].

We studied CIMS in the same devices and compared the results with those from FIMS. We held the $Ni_{0.8}Fe_{0.2}$ magnetization fixed by applying a magnetic field of a magnitude between the switching fields of $Ni_{0.8}Fe_{0.2}$ and Ni and then applied a current pulse to switch the Ni magnetization. (The applied field changes the $Ni_{0.8}Fe_{0.2}$ magnetic energy landscape from bi-stable to mono-stable, effectively fixing the magnetization in one direction.) If the bias current density exceeds $\approx 5 \times 10^6$ A/cm$^2$, the device resistance increases by a factor of two or more because the Nb electrodes in the nanopillar become resistive[30]. This also results in a change in $I_{cs}$ due to trapped magnetic flux, which we removed by briefly heating the chip above $T_{cs}$ before measuring each $I_{cs}$; see Fig. 2a for the control pulse sequence. Fig. 2b shows hysteretic switching of $I_{cs}$ to high or low values depending on the current pulse polarity. Positive current is associated with electron flow from Ni to $Ni_{0.8}Fe_{0.2}$ (Fig. 1a). Switching to an AP (or P) state with a positive (or negative) current is a signature of the standard STT effect. This asymmetry rules out the Oersted field effect as the prevailing factor in this CIMS[26,31]. The P and AP states are reached at $3 \times 10^7$ A/cm$^2$ and $5 \times 10^7$ A/cm$^2$, respectively, which are of the same order of magnitude as the switching current density $J_{cm}$ found in comparable studies on room-temperature devices[26,31,32] but higher than the maximum supercurrent density ($\approx 5 \times 10^6$ A/cm$^2$) of the Nb electrodes in the nanopillars. CIMS consists of multiple jumps during the transitions (Fig. 2b) as in the FIMS of the Ni magnetization (Fig. 1d). These multiple jumps may indicate the presence of magnetic nano-domains.

We also obtained FIMS loops that can be used to determine the Ni magnetization orientation after CIMS. Fig. 2c shows a loop obtained after both the $Ni_{0.8}Fe_{0.2}$ and Ni magnetizations are saturated with a high field. For CIMS, we applied a +5 mA current pulse (as well as a field that holds the $Ni_{0.8}Fe_{0.2}$ magnetization direction only) after such a saturating field (Fig. 2d inset). An FIMS loop measured subsequently has the reversed symmetry indicating a reversed (negative) Ni magnetization (Fig. 2d). With consecutive +5 mA and −5 mA current pulses after a saturating field, we obtain a positive Ni magnetization (through two magnetization reversals) and confirm that a negative current pulse also switches the Ni magnetization (Fig. 2e).

According to the standard STT theory, the switching current threshold increases with $M_sV$ of the free layer, where $M_s$ and $V$ are the saturation magnetization and volume, respectively, if other parameters are fixed[25,27]. Since this suggests that the same magnetic materials of different thicknesses may be used to obtain CIMS in the resulting S-PSV-S JJs, we developed nanopillar JJs with a Ni(1)/Cu/Ni(2.4)-based PSV barrier. Without $Ni_{0.8}Fe_{0.2}$, there is a smaller number of materials parameters for analysis and the reduced electron scattering associated with a non-alloyed material results in less supercurrent decay, enabling us to explore a higher $J_{cs}$ regime. The switching field ranges of Ni(1) and Ni(2.4) layers are not well-separated from each other (as confirmed with magnetization measurements on unpatterned Ni films), which limits the control of the PSV magnetization state with field between P and partially switched states. Fig. 3a shows an $I_{cs}$ vs. field pulse height characteristic measured the same way as the case of Fig. 1d. Non-P states result in $I_{cs} < I_{cs}^P$, which indicates $I_{cs}^{AP}$ is also lower than $I_{cs}^P$ similar to the $Ni_{0.8}Fe_{0.2}$(0.8 or 1)/Cu/Ni(2.4)-based devices as expected.

Figures 3b–3f show the hysteresis loops in the measured $I_{cs}$ vs. current-pulse height without an applied magnetic field. Switching to a P (or AP) state with a positive (or negative) current is consistent with the switching of the lower (thin) Ni relative to the upper Ni through the standard STT effect. The switching current increases with area (or total magnetic moment) as expected. P and AP states are reached at current densities slightly higher than those of $Ni_{0.8}Fe_{0.2}$/Cu/Ni devices. There are fewer or no intermediate states in smaller devices, which indicates they are more nearly single-domain, approaching a two-state regime. Comparing Fig. 3a and 3b, we find that CIMS results in the same maximum $I_{cs}$ (resulting from the same P state) and a lower minimum $I_{cs}$ compared with FIMS. Although this alone does not confirm the minimum $I_{cs}$ obtained with CIMS is from the AP state, the maximum and minimum $I_{cs}$ values have about the same ratio ($\approx$ 3:1) in different devices (Figs. 3b–3f) and each also scales with area without significant scatter (Fig. 3g), which suggest that the maximum and minimum $I_{cs}$s are likely to be associated with well-defined P and AP states instead of intermediate states. Fig. 3h shows the mean $I_{cs}R_n$ of the devices presented in Fig. 3g together with the calculated $I_{cs}R_n$ vs. the thickness of the hard layer Ni characteristics obtained by use of the fitted material parameters given in Ref. 10 (except the prefactor). The calculation predicts the correct sign in $\Delta I_{cs}$. The slight underestimation in the $I_{cs}$ ratio may be due to the uncertainty in estimating the magnetic layer thicknesses (or the effective magnetic dead layer thicknesses[10,33]). The calculation suggests that both states are in the $\pi$-JJ regime and, with the upper Ni layer thickness $\approx$ 1.9 nm, we could obtain 0-$\pi$ phase-switching devices that are controlled with the STT effect[7,10,34].

Only a few theoretical studies discuss the impact of superconducting electrodes on STT[35,36,37,38,39]. The major difference between a superconducting and a non-superconducting system is the presence of Andreev reflections below the superconducting gap voltage at the superconductor-normal metal interfaces, resulting in zero spin current[35]. In our devices, the contribution from the Andreev reflections should not be significant and the STT effect should be similar to a non-superconducting case, since the CIMS occurs at higher voltages ($\approx$20 mV) than $2\Delta_{Nb}$ ($\approx$3 mV at the bulk limit where $\Delta_{Nb}$ is the Nb gap voltage). In future work, $J_{cm}$ may be reduced by appropriately engineering the magnetic materials. This may make the STT effect practical for high-density superconducting memory applications and also allow observation of an STT effect that is significantly different from a non-superconducting case.

Fundamentally, how small a memory element can be made is limited by its thermal stability and the required data retention time. Using our experimental results, we estimate the FIMS magnetic energy barrier is on the order of $10^{-20}$ J and $10^{-19}$ J for a 1 nm thick, 50 nm × 100 nm elliptical $Ni_{0.8}Fe_{0.2}$ and Ni, respectively. This is well above $60k_BT = 3 \times 10^{-21}$ J at 4 K ($k_B$ is the Boltzmann constant), the energy barrier commonly required for long-term memory stability, and our results suggest that even smaller devices with lower switching energies are possible[40]. On the other hand, JJs for superconducting digital electronics are commonly designed to have $I_{cs}$ of at least $\approx$100 μA to make the Josephson energy $E_J = I_{cs}\Phi_0/2\pi$ much larger than $k_BT$[11], which is a more stringent requirement. However, a lower $I_{cs}$ and $E_J$ may be allowed for cryogenic memory elements because retention is determined by the magnetic properties of the PSV, while the Josephson effect could be considered a function of a magnetic-to-electrical transducer and needs to be stable for only the short duration of a memory read operation.

**Acknowledgments**

This work was supported by NIST and by US National Security Agency under agreement numbers EAO156513 and EAO176792.

# References

[1] Buzdin, A. I., Bulaevskij, L. N. & Panyukov, S. V. Critical-current oscillations as a function of the exchange field and thickness of the ferromagnetic metal (F) in an S-F-S Josephson junction. *J. Exp. Theor. Phys. Lett.* **35**, 178–180 (1982).

[2] Ryazanov, V. V. *et al*. Coupling of Two Superconductors through a Ferromagnet: Evidence for a π Junction, *Phys. Rev. Lett.* 86, 2427-2430 (2001).

[3] Kontos, T. *et al.* Josephson junction through a thin ferromagnetic layer: negative coupling. *Phys. Rev. Lett.* **89**, 137007 (2002).

[4] Buzdin, A. I. Proximity effects in superconductorferromagnet heterostructures. *Rev. Mod. Phys.* **77**, 935976 (2005).

[5] Bergeret, F. S., Volkov, A. F. & Efetov, K. B. Odd triplet superconductivity and related phenomena in superconductor-ferromagnet structures. *Rev. Mod. Phys.* **77**, 1321–1373(2005).

[6] Feofanov, A. K. *et al.* Implementation of superconductor/ferromagnet/superconductor π-shifters in superconducting digital and quantum circuits. *Nat. Phys.* **6**, 593–597 (2010).

[7] Bell, C. *et al.* Controllable Josephson current through a pseudospin-valve structure. *Appl. Phys. Lett.* **84**, 1153–1155 (2004).

[8] Goldobin, E. *et al.* Memory cell based on a φ Josephson junction. *Appl. Phys. Lett.* **102**, 242602 (2013).

[9] Qader, M. A. E. et al. Switching at small magnetic fields in Josephson junctions fabricated with ferromagnetic barrier layers, *Appl. Phys. Lett.* **104**, 022602 (2014).

[10] Baek, B. *et al.* Hybrid superconducting-magnetic memory device using competing order parameters. *Nat. Commun.* **5**, 3888 (2014).

[11] Likharev, K. K. & Semenov, V. K. RSFQ logic/memory family: a new Josephson-junction technology for sub-terahertz-clock-frequency digital systems. *IEEE Trans. Appl. Supercond.* **1**, 328 (1991).

[12] Bedard, F., Welker, N., Cotter, G. R., Escavage, M. A. & Pinkston, J. T. *Superconducting Technology Assessment* Nat. Security Agency Office Corp. Assessments: Fort Meade, MD, USA, (2005).

[13] Mukhanov, O. A. Energy-Efficient Single Flux Quantum Technology. *IEEE Trans. Appl. Supercond.* **21**, 760–769 (2011).

[14] Holmes, D. S., Ripple, A. L. & Manheimer, M. A. Energy-efficient superconducting computing—power budgets and requirements. *IEEE Trans. Appl. Supercond.* **23**, 1701610 (2013).

[15] Larkin, T. *et al.* Ferromagnetic Josephson switching device with high characteristic voltage. *Appl. Phys. Lett.* **100**, 222601 (2012).

[16] Herr, A. Y. & Herr, Q. P. Josephson magnetic random access memory system and method. US Patent 8,270,209 B2 filed 30 April 2010, and issued 18 September 2012.

[17] Park, J., Ralph, D. C. & Buhrman, R. A. Fast deterministic switching in orthogonal spin torque devices via the control of the relative spin polarizations. *Appl. Phys. Lett.* **103**, 252406 (2013).

[18] Ye, L. *et al.* Spin-transfer switching of orthogonal spin-valve devices at cryogenic temperatures. *Appl. Phys. Lett.* **115**, 17C725 (2014).

[19] Niedzielski, B. M. *et al.* Use of Pd–Fe and Ni–Fe–Nb as Soft Magnetic Layers in Ferromagnetic Josephson Junctions for Nonvolatile Cryogenic Memory. *IEEE Trans. Appl. Supercon.* **24**, 1800307 (2014).

[20] Bergeret, F. S., Volkov, A. F. & Efetov, K. B. Enhancement of the Josephson current by an exchange field in superconductor-ferromagnet structures. *Phys. Rev. B* **64**, 134506 (2001).

[21] Blanter, Ya. M. & Hekking, F. W. J. Supercurrent in long SFFS junctions with antiparallel domain configuration. *Phys. Rev. B* **69**, 024525 (2004).

[22] Linder, J. & Sudbø, A. Josephson effect in thin-film superconductor/insulator/superconductor junctions with misaligned in-plane magnetic fields. *Phys. Rev. B.* **76**, 064524 (2007).

[23] Demler, E. A., Arnold, G. B. & Beasley, M. R. Superconducting proximity effects in magnetic metals. *Phys. Rev. B* **55**, 15174–15182 (1997).

[24] Eschrig, M. Spin-polarized supercurrents for spintronics. *Phys. Today* **64**, 43–49 (2011).
[25] Slonczewski, J. C. Current-driven excitation of magnetic multilayers. *J. Magn. Magn. Mater.* **159**, L1 (1996).
[26] Wolf, S. A. *et al.* Spintronics: A Spin-Based Electronics Vision for the Future *Science*. **294**, 1488–1495 (2001).
[27] Katine, J. A., Albert, F. J., Buhrman, R. A., Myers, E. B. & Ralph, D. C. Current-driven magnetization reversal and spin-wave excitations in Co/Cu/Co pillars. *Phys. Rev. Lett.* **84**, 3149–3152 (2000).
[28] Tinkham, M. *Introduction to Superconductivity* McGraw-Hill: New York, (1975).
[29] Larkin, T. *et al.* Ferromagnetic Josephson switching device with high characteristic voltage. *Appl. Phys. Lett.* **100**, 222601 (2012).
[30] Huebener, R. P., Kampwirth, R. T., Martin, R. L., Barbee, Jr., T. W. & Zubeek, R. B., Critical current density in superconducting niobium films. *J. Low Temp. Phys.* **19**, 247–258 (1975).
[31] Grollier, J. *et al*. Spin-polarized current induced switching in Co/Cu/Co pillars. *Appl. Phys. Lett.* **78**, 3663–3665 (2001).
[32] Lacour, D., Katine, J. A., Smith, N., Carey, M. J. & Childress, J. R. Thermal effects on the magnetic-field dependence of spin-transfer-induced magnetization reversal. *Appl. Phys. Lett.* **80**, 4681–4683 (2004).
[33] Speriosu, V. S. *et al.* Role of interfacial mixing in giant magnetoresistance. *Phys. Rev. B* **47**, 11579–11582 (1993).
[34] Mielke, O., Ortlepp, Th., Dimov, B. & Uhlmann, F. H. Phase engineering techniques in superconducting quantum electronics. *J. Phys. Conf. Ser.* **97**, 012196 (2008).
[35] Waintal, X. & Brouwer, P. W. Current-induced switching of magnetic domains to a perpendicular configuration. *Phys. Rev. B*. **63**, 220407(R) (2001).
[36] Waintal, X. & Brouwer, P. W. Magnetic exchange interaction induced by a Josephson current. *Phys. Rev. B*. **65**, 054407 (2002).
[37] Zhao, E. & Sauls, J. A. Theory of nonequilibrium spin transport and spin-transfer torque in superconducting-ferromagnetic nanostructures. *Phys. Rev. B* **78**, 174511 (2008).
[38] Linder, J. and Yokoyama, T. Supercurrent-induced magnetization dynamics in a Josephson junction with two misaligned ferromagnetic layers. *Phys. Rev. B* **83**, 012501 (2011).
[39] Hoffman, S., Blanter, Y. M. & Tserkovnyak, Y. Nonlinear dynamics in a magnetic Josephson junction. *Phys. Rev. B* **86**, 054427 (2012).
[40] Gajek, M. *et al.* Spin torque switching of 20nm magnetic tunnel junctions with perpendicular anisotropy. *Appl. Phys. Lett.* **100**, 132408 (2012).

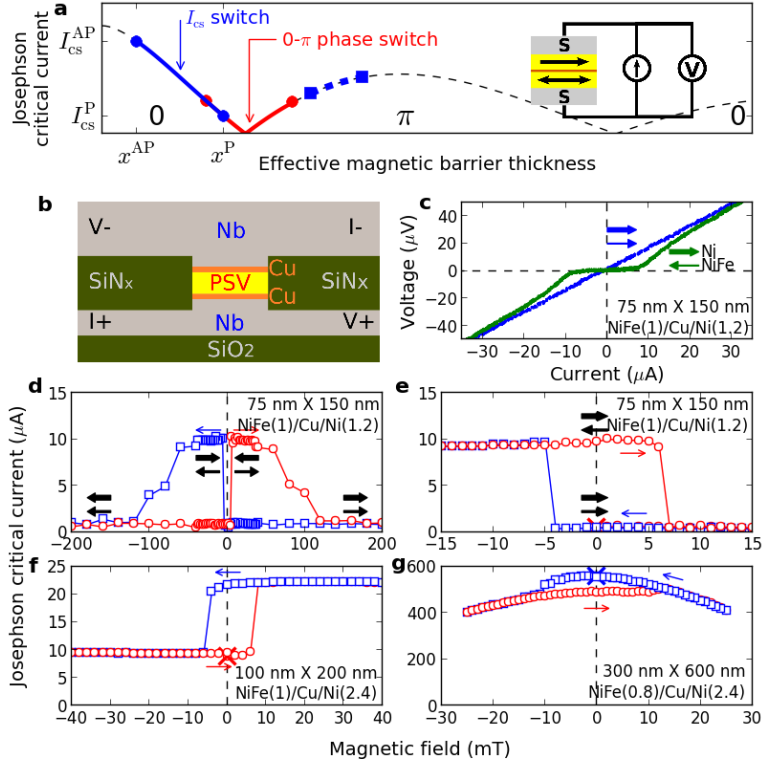

Fig. 1. Nanopillar JJs with a PSV-barrier $Ni_{0.8}Fe_{0.2}/Cu/Ni$. (a) Illustrated oscillation in Josephson critical current ($I_{cs}$) with effective magnetic barrier thickness $x$. Changes in $I_{cs}$ (blue) or phase (red) in S-PSV-S JJs of different magnetic barrier thicknesses are shown with thick curves. For simplicity carrier scattering, noncollinear magnetization changes, and domain structure effects are not considered. Inset: PSV-barrier JJ model and measurement circuit. (b) Device structure and measurement lead configuration. (c) Voltage vs. current characteristics at zero applied field. (d) Wide-range hysteresis loop of $I_{cs}$ vs. magnetic field pulse height. Each $I_{cs}$ was measured at zero applied field after we applied the magnetic field pulse followed by a heat pulse. (e–g) Minor magnetic hysteresis loops of $I_{cs}$ for different devices. Ni moment was set to the positive maximum with 400 mT before each field sweep. Data in (c–e) were obtained from the same device. The dimensions in the figure represent minor and major axes of an elliptical device design. The field direction is parallel to the major axis of the device. Red circles and blue squares are for upward and downward field (or field pulse height) sweeps, respectively, and this differentiation of swept

field or current directions is applied to other figures as well. X represents the start of the sweep. A thick and thin arrow pair indicates the Ni and $Ni_{0.8}Fe_{0.2}$ magnetization directions, respectively.

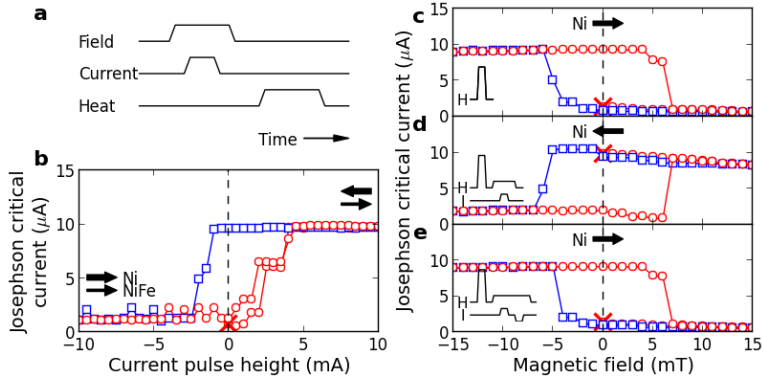

Fig. 2. Current-induced PSV magnetization switching in a nanopillar JJ. The PSV structure is $Ni_{0.8}Fe_{0.2}$ (1)/Cu/Ni(1.2). The elliptical device design dimensions are 75 nm × 150 nm (the same device as for Figs. 1c–1e). (a) Control pulse (on-off) sequence applied before measuring each $I_{cs}$. The pulse durations and delays are 1 s – 5 s. (b) Hysteretic $I_{cs}$ vs. current pulse height obtained with the sequence in (a) before measuring each $I_{cs}$. The initial P state (marked with X) is preset with a 400 mT field. The field pulse height is 15 mT for every datum. Arrows indicate the inferred PSV magnetization state based on $I_{cs}$ (low $I_{cs}$: P vs. high $I_{cs}$: AP). (c–e) Minor magnetic hysteresis loops (showing $Ni_{0.8}Fe_{0.2}$ layer switching) after an applied control pulse sequence as illustrated in each inset. Inset: The first high $H$ pulse represents the 400 mT field pulse while the second lower $H$ pulse is 15 mT. The positive and negative $I$ pulses represent +5 mA and −5 mA current pulses, respectively. Thick and thin arrow pairs indicate the Ni and $Ni_{0.8}Fe_{0.2}$ magnetization directions. A single thick arrow represents the Ni magnetic moment direction inferred from the minor hysteresis loop shape.

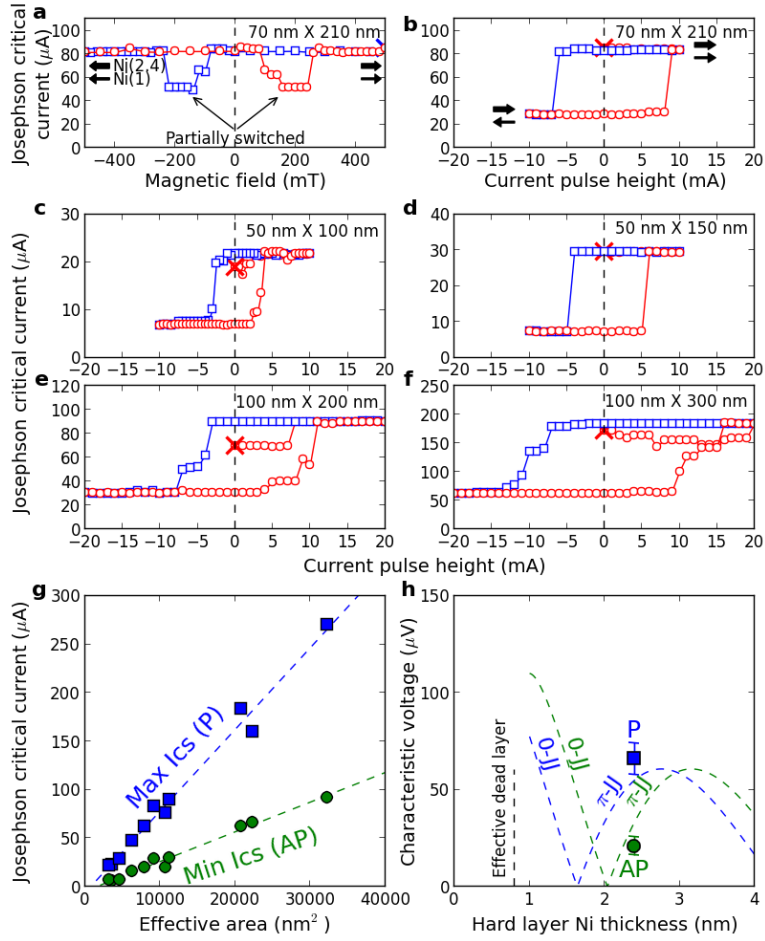

Fig. 3. PSV magnetization switching in a nanopillar JJ with an all-Ni PSV barrier. The PSV structure is Ni(1)/Cu/Ni(2.4). (a) Wide-range hysteresis loop of $I_{cs}$ vs. magnetic field pulse height. A thin and thick arrow pair indicates the Ni(1) and Ni(2.4) magnetization directions. (b–f) Hysteretic $I_{cs}$ vs. current pulse height. Data in (b) was obtained from the same device for (a). (g) Maximum and minimum $I_{cs}$ vs. effective device area $A_{\text{eff}}$ of devices on the same chip. Each $A_{\text{eff}}$ was estimated by linear-fitted $R_n A$ (vs. $A$) divided by $R_n$. Dashed lines are linear fits. (h) Averaged Josephson characteristic voltages $I_{cs}R_n$'s (symbols) from (g) plotted together with calculated curves (dashed curves), which were calculated with the same fitting parameters (characteristic oscillation length 1.0 nm and Ni dead layer thickness $d_{\text{dead}} =$ 0.8 nm) obtained in Ref. 10. Each error bar represents a standard error of the mean.